\documentclass[aps,prl,reprint,superscriptaddress]{revtex4-1}

\usepackage{physics}
\usepackage{graphicx}
\usepackage{amsmath}
\usepackage{color}
\usepackage{multirow}
\graphicspath{{images/}}

\begin{document}

\title{Switching a polar metal via strain gradients}

\author{Asier Zabalo} 
\affiliation{Institut de Ci\`encia de Materials de Barcelona (ICMAB-CSIC), Campus UAB, 08193 Bellaterra, Spain}
\author{Massimiliano Stengel}
\affiliation{Institut de Ci\`encia de Materials de Barcelona (ICMAB-CSIC), Campus UAB, 08193 Bellaterra, Spain}
\affiliation{ICREA-Instituci\'o Catalana de Recerca i Estudis Avançats, 08010 Barcelona, Spain}

\date{\today}

\begin{abstract}
Although rare, spontaneous breakdown of inversion symmetry sometimes occurs in
a material which is metallic: these are commonly known as polar metals or ferroelectric metals.
Their \textit{polarization}, however, cannot be switched via an electric field, which limits 
the experimental control over band topology. Here we shall investigate, via first-principles 
theory, flexoelectricity as a possible way around this obstacle with the well known polar 
metal $\text{LiOsO}_3$. 
The flexocoupling coefficients are computed for this metal with high accuracy with a 
completely new approach based on real-space sums of the inter-atomic 
force constants. 
A Landau-Ginzburg-Devonshire-type first-principles Hamiltonian is built and a 
critical bending radius to switch the material is estimated, whose order of magnitude
is comparable to that of BaTiO$_3$. 
\end{abstract}

\maketitle
The so-called ‘polar’ or ‘ferroelectric’ 
metals \cite{kim2016polar}, first proposed by Anderson more than half a
century ago in the context of martensitic transformations \cite{anderson1965symmetry},
have been attracting increasing attention recently. 
The prototypical (and historically the first experimentally known) material realization is lithium osmate, which undergoes a ferroelectric-like transition at 140 K from the 
centrosymmetric $R\bar{3}c$ to the non-centrosymmetric $R3c$ space 
group \cite{shi2013ferroelectric}. 
Since its discovery, the list of known polar metals has been steadily growing \cite{benedek2016ferroelectric}. 
Their interest lies on the unusual physics that may emerge from
the coexistence of metallicity and polarity, two properties that
were initially regarded as contraindicated.
For instance, they provide excellent opportunities to study exotic quantum phenomena, 
like non-centrosymmetric superconductivity \cite{bauer2012non,yip2014noncentrosymmetric} 
or spin-polarized currents \cite{lu2010spin}.
In spite of considerable progress, however, a long-standing issue still remains, 
and concerns the ability to control polarity via an appropriate external field.
Indeed, due to the presence of free carriers in the bulk the most obvious means of
switching polarity in ferroelectrics, i.e. an external electric field, is ruled out.
Such a control would help shed some light on  
their fundamental physics, and possibly devise some applications, e.g., 
in nanoscale electronic and thermoelectric devices \cite{ma2018design}.
Our goal is to demonstrate that flexoelectricity can solve this issue.  
  
Flexoelectricity describes the coupling between a strain gradient and the macroscopic 
polarization and, unlike its homogeneous counterpart (piezoelectricity), it does not 
require any particular space group to be 
present \cite{stengel2016flexoelectricity,yudin2013fundamentals,stengel2013flexoelectricity}. 
While flexoelectricity is hardly a new discovery~\cite{kogan1964piezoelectric}, its 
practical relevance was demonstrated only recently, thus reviving this field 
from both the experimental \cite{vasquez2018flexoelectricity,narvaez2016enhanced,lu2012mechanical} 
and theoretical \cite{royo2019first} points of view.
Of course, the electrical polarization can only be defined in insulating crystals,
so the macroscopic flexoelectric coefficient of a polar metal vanishes identically.
Hovever, as we shall demonstrate shortly, the coupling between polar lattice 
modes and a strain gradient does exist even in metals.
Since elastic fields, unlike electric fields, are not screened by free carriers, this
constitutes, in principle, a viable means of controlling polarity.
Still, whether the relevant couplings are strong enough for such a mechanism to be experimentally 
accessible, is currently unknown. 
First-principles calculations could be very helpful in this context,
and indeed electronic-structure methods to study flexoelectricity have seen an impressive 
progress in recent years \cite{stengel2017first,royo2019first}. 
However, the calculation of the flexocoupling coefficients remains a subtle task even with insulators and treating metallic crystals falls outside the present capabilities of the density functional theory (DFT) based codes. 

Here we overcome such limitations by developing an accurate and general
method to calculate flexocoupling coefficients in metals, which is
based on real space sums of the inter-atomic force constants (IFC-s).
We demonstrate our computational strategy by calculating the flexocoupling coefficients for $\text{LiOsO}_3$ as a test case and 
we compare them with the ones of $\text{BaTiO}_3$, probably one of the most studied ferroelectric 
materials.
Finally, we use the aforementioned values, in combination with a first-principles based effective Hamiltonian
that we have constructed by expanding the energy around the centrosymmetric cubic phase, to estimate
the critical bending radius of $\text{LiOsO}_3$. 
We find the values in line with those calculated for $\text{BaTiO}_3$, a material where flexoelectric
switching of the polar domains has been experimentally demonstrated already.
Based on these results, mechanical switching of LiOsO$_3$ mediated by flexoelectricity appears well
within experimental reach.
\begin{figure}
	\includegraphics[width=1.0\linewidth]{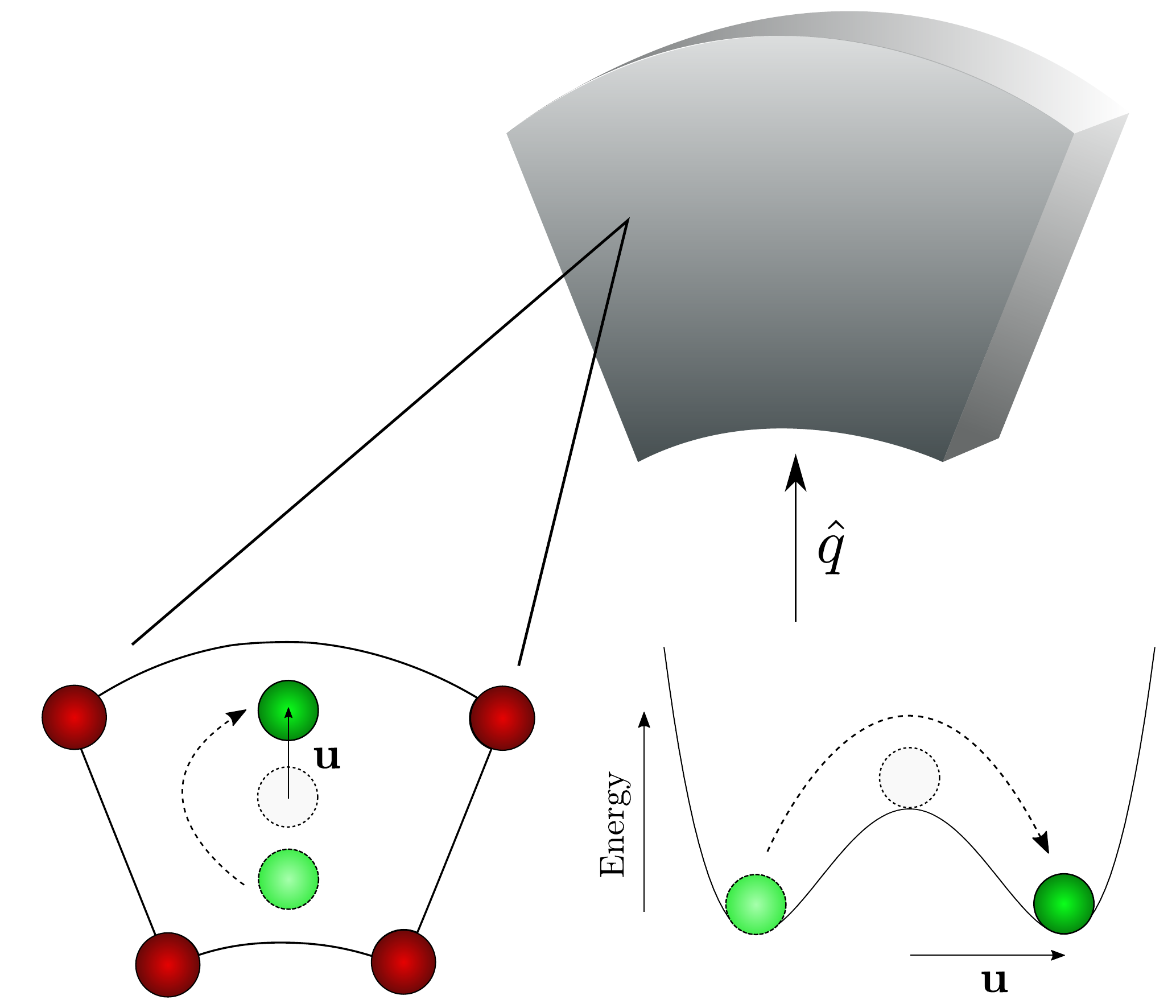}
	\caption{A bending type strain gradient is applied to a macroscopic crystal along the direction $\hat{q}$. The external strain gradient couples to the polar modes resulting in a displacement of the atoms and, as a consequence, the structure evolves to another symmetrically equivalent ferroelectric state.}
	\label{fig:strain_gradient}
\end{figure}

To start with, we shall consider a setup as illustrated in Fig. 1,
i.e. of a LiOsO$_3$ (or $n$-doped BaTiO$_3$) sample that is cut along some 
crystallographic direction $\hat{q}$ and mechanically bent via some external load.
Within the interior of the film, the polar order parameter is assumed to be
homogeneous, and its amplitude is described by some three-dimensional vector ${\bf u}$
with the physical dimension of length.
In the following we shall quantify the \emph{coercive
bending radius}, i.e. the radius of curvature that needs to be applied
in order to switch the polar order parameter between two neighboring local
minima, which are degenerate at mechanical equilibrium.
We shall calculate the critical radius via the following formula,
\begin{equation}\label{eq:R_crit}
R_{\rm crit} = \frac{f_{\rm eff}}{F_{\rm coerc}},
\end{equation}
where $f_{\rm eff}$
is the effective \emph{flexocoupling}
coefficient associated with the flexural deformation, and $F_{\rm coerc}$
is the minimal generalized force that is required for the mode ${\bf u}$
to cross the energy barrier between two minima. 
Thus, the problem can be divided into two separate tasks: (i) determining
the coupling between a flexural deformation and the polar mode, described
by $f_{\rm eff}$, as a function of the crystallographic orientation, and 
(ii) identifying the most likely switching paths and the corresponding 
energetics.

From now on, we shall assume a Landau-like expansion of the energy
around the high-symmetry cubic structure as a function of the relevant
parameters, following the established common practice in theoretical
studies of perovskite ferroelectrics.
In this context, task (ii) entails no conceptual difficulties, as it 
consists in mapping the potential energy surface of the crystal as a 
function of the relevant lattice degrees of freedom -- such a procedure
has been successfully carried out for a wide range of materials already.
The main technical obstacle resides in (i), since no established 
methods exist for the calculation of $f_{\rm eff}$ in metals. Given
the novelty, we shall focus on this point in the following.

In full generality, the flexocoupling tensor in a ``soft-mode'' material 
can be defined as follows \cite{stengel2016unified},
\begin{equation}\label{eq:flexo_ifc}
f^{\rm I}_{\alpha \beta, \gamma \lambda}= 
    - \frac{M}{2} \bra{P_\alpha}  
   \frac{\partial^2 D({\bf q}) }{\partial q_\gamma \partial q_\lambda} \bigg|_{{\bf q}=0} \ket{A_\beta} 
\end{equation}
where $D({\bf q})$ is the dynamical matrix of the crystal at a certain wavevector ${\bf q}$,
and the bra and kets represent the TO$_1$ and acoustic eigenvectors at the $\Gamma$ ($\mathbf{q}=0$)
point of the Brillouin zone. 
This tensor describes the force on $u_\alpha$ that is produced by a macroscopic
strain gradient $\eta_{\beta,\gamma\lambda}$, the latter expressed in ``type-I'' form
(hence the superscript ``I''), i.e., as the second gradient of the displacement field. 
For our present scopes, it is more convenient to work in type-II form,
which can be recovered via the following transformation,
\begin{equation}\label{Eq:type_relation}
f^{\rm II}_{\alpha\lambda,\beta\gamma}=f^{\rm I}_{\alpha\beta,\gamma\lambda}+f^{I}_{\alpha\gamma,\lambda\beta}-f^{\rm I}_{\alpha\lambda,\beta\gamma}.
\end{equation}
The main technical challenge from a computational point of view consists in taking 
the second gradient with respect to ${\bf q}$ of the
dynamical matrix, $D({\bf q})$.  
In insulators, this task is already delicate at the formal level, since $D({\bf q})$ has a \emph{nonanalytic} behavior in vicinity of $\Gamma$; this
requires a careful treatment of the macroscopic electric fields before
performing the perturbative long-wave expansion.
In metals \emph{at finite temperature} things appear simpler conceptually, since the adiabatic dynamical matrix $D({\bf q})$ is an analytic function of the wavevector ${\bf q}$ over the whole Brillouin zone. This means that the second ${\bf q}$-gradient appearing in Eq.~(\ref{eq:flexo_ifc}) is always well defined without taking any further precaution. 
However, this methodology has not been generalized to metals yet. To circumvent this obstacle,
we define the long-wave expansion of the dynamical matrix as the real-space moments 
of the interatomic force constants, following the method described in 
\cite{stengel2013flexoelectricity,stengel2016unified}. 
In particular, the IFC's are first defined as the second derivative of the total energy with
respect to atomic displacements,
\begin{equation}
\Phi^l_{\kappa\alpha,\kappa'\beta}=\frac{\partial^2 E}{\partial u_{\kappa\alpha}^{0}\partial u_{\kappa'\beta}^{l}}.
 \label{ifc}
\end{equation}
Then, we write
\begin{equation}
\frac{\partial^2 D^{({\bf q})}_{\kappa \alpha, \kappa'\beta} }
     {\partial q_\gamma \partial q_\lambda} \bigg|_{{\bf q}=0} = 
   \sum_l \frac{\Phi^l_{\kappa\alpha,\kappa'\beta}}{\sqrt{m_\kappa m_{\kappa'}}}
   (\mathbf{d}^l_{\kappa \kappa'})_\gamma (\mathbf{d}^l_{\kappa \kappa'})_\lambda,
   \label{d2sum}
\end{equation}
where $m_\kappa$ is the mass of atom $\kappa$,
$\mathbf{d}^l_{\kappa \kappa'}= \mathbf{R}^l+\boldsymbol{\tau}_{\kappa'}-\boldsymbol{\tau}_\kappa$, 
$\mathbf{R}^l$ is the Bravais lattice vector indicating the location of the 
$l$-th cell and $\boldsymbol{\tau}_\kappa$ is the position of atom $\kappa$ within the unit 
cell $l=0$.
The short-range nature of the interatomic forces guarantees that
the lattice sums of Eq.~(\ref{d2sum}) will eventually converge to the 
correct physical value when a dense enough ${\bf q}$-point mesh is used 
to calculate the real-space force constants of Eq.~(\ref{ifc}). Interestingly, the coupling between two acoustic modes
is directly related (in a crystal that is free of stresses) 
to the elastic tensor components via \cite{born1954dynamical} 
\begin{equation}\label{eq:elastic_ifc}
\frac{ C_{\alpha \gamma,\beta \lambda} + C_{\alpha \lambda,\beta \gamma } }{2} = 
    - \frac{M}{2\Omega} \bra{A_\alpha}  
   \frac{\partial^2 D({\bf q}) }{\partial q_\gamma \partial q_\lambda} \bigg|_{{\bf q}=0} \ket{A_\beta}.
\end{equation}
This is a useful consistency check: one can then compare the results with a more conventional calculation of the elastic 
tensor \cite{wu2005systematic} to gauge the reliability of the flexocoupling coefficients as determined via Eq.~(\ref{eq:flexo_ifc}).
Note that the elastic tensor components are themselves a crucial ingredient for 
calculating the effective flexocoupling of Eq. (\ref{eq:R_crit})  starting from the flexocoupling tensor 
${\bf f}^{\rm II}$; therefore, it is important to ensure that the two physical quantities
are calculated with consistent accuracy. 
\begin{table}
	\caption{Independent components of the elastic (in GPa) and flexocoupling (in eV) tensor with a $16\cross16\cross1$ mesh of \textbf{q} points for the IFC-s. The n-type flexocoupling coefficients of $\text{BaTiO}_3$ are shown here. \label{tab:elastic_comparsion}}
	\begin{ruledtabular}
		\begin{tabular}{cccccccc}
			&& $C_{11}$& $C_{12}$ & $C_{44}$ &$f^{\rm II}_{11}$& $f^{\rm II}_{12}$ & $f^{II}_{44}$  \\
			\hline
			\multirow{2}*{$\text{LiOsO}_3$}&Long-wave & 364.7 & 129.5 & 44.3 & $-$13.8 & 49.3 & 3.3\\
			&DFPT& 365.6 & 129.5 & 44.1 &---&---&---\\
			\hline
      \multirow{2}*{$\text{BaTiO}_3$}& Long-wave & 346.1 & 121.7 & 134.5 & $-$53.5&3.4  & $-$39.5\\
      &DFPT & 353.3 & 121.7 & 137.7  &---&---&---\\
		\end{tabular}
	\end{ruledtabular}
\end{table}   

Our first principles calculations are performed with the open-source \textsc{abinit} \cite{gonze2009abinit,gonze2020abinit} package. (Details of the computational parameters are provided in the Supplementary Material.) Numerical results for both $\text{BaTiO}_3$ and LiOsO$_3$ are shown in Table \ref{tab:elastic_comparsion}. 
Clearly, the largest flexocouplings are $f_{12}$ in
LiOsO$_3$ and $f_{11}$ for BaTiO$_3$. (The latter material behaves very similarly to SrTiO$_3$ 
\cite{stengel2016unified},
which is natural to expect given the affinities in the electronic and atomic structure.)
Their absolute values are similar overall, which provides a first indication that the 
flexocoupling is comparably strong in these two materials. 
Note that the discrepancy in the elastic constants calculated via the two different methods
is less than a 1 \% for the three independent components of LiOsO$_3$, which confirms the excellent 
quality of the calculations.  
We also show in Fig. S3 (and Table S4) the convergence of the numerical results for both the
elastic and flexocoupling constants as a function of the ${\bf q}$-point mesh, 
further corroborating this point.
 
To make further progress, we use the value of Table \ref{tab:elastic_comparsion} to compute 
the effective flexocoupling coefficients 
for three representative orientations of the sample ([100], [110] and [111]), either in 
the beam-bending or the plate-bending limit.
(We focus on the beam-bending limit following the definitions of Ref. \cite{narvaez2015large}; 
explicit formulas are reported in the Supplementary Material.)
The results, shown in Table \ref{tab:bending_radius}, indicate that 
[100] is by far the bending direction that produces the largest flexocoupling
in LiOsO$_3$.
The situation in BaTiO$_3$ seems to be more balanced overall, with a slight 
preference for [110] and [111] directions over [100].
Note, however, that for each surface orientation $\hat{\bf q}$, the effective 
flexocoupling describes the flexo-induced force acting on the polar mode along $\hat{\bf q}$.
Depending on the switching path, such force might not be
parallel to the direction along which the polar mode evolves during switching, $\hat{\bf s}$;
in such cases the effective flexocoupling needs to be scaled by the projection 
$\hat{\bf q} \cdot \hat{\bf s}$.
Since the relevant paths in BaTiO$_3$ (see next paragraph) involve [100]-oriented
switching, such geometrical factor reduces the [110]  and [111] coefficients by $\sqrt{2}$ and $\sqrt{3}$ respectively, bringing all three values of $f_{\rm eff}$ to a similar magnitude.
\begin{table}
  \caption{Effective flexocoupling coefficients (absolute values) for 100, 110 and 111 oriented samples (in eV units) for the beam-bending limit. 
  \label{tab:bending_radius}}
  \begin{ruledtabular}
    \begin{tabular}{cccc}
      &$f_{\text{eff}}^{100}$&$f_{\text{eff}}^{110}$ & $f_{\text{eff}}^{111}$ \\ 
      $\text{LiOsO}_3$& 40.1 &  5.3  & 2.5 \\
      $\text{BaTiO}_3$& 16.2 & 24.7  &23.3
    \end{tabular}
  \end{ruledtabular}
\end{table}

Having calculated the values of $f_{\rm eff}$, we now need the 
information about the switching path to obtain $R_{\rm crit}$ 
according to Eq. (\ref{eq:R_crit}).
To this end, we construct a Landau-Ginzburg-Devonshire-type first-principles 
Hamiltonian by expanding the energy around the reference cubic phase,
of $Pm\bar{3}m$ symmetry.
The Hamiltonian includes the most important degrees of freedom of the structure: 
the strain $s_{ij}$, the tilts of the oxygen octahedra $q_i$, where $q_i$ represents 
the displacements of the oxygen atoms perpendicular to the rotation axis and the polar modes $u_i$. 
(Details on the model can be found in the Supplementary Material.)
First, we validate our effective Hamiltonian $H_{\rm eff}$ by calculating
the energetics of the relevant phases (Table S2) and their variation as
a function of external pressure (Fig. S1); in both cases we obtain excellent
agreement to the first-principles results. Next, we proceed to calculating the most favorable switching paths 
by constraining one component of the polar vector~\cite{dieguez2006first,stengel2009electric}
and numerically minimizing (simulated annealing) the energy functional 
with respect to the other parameters.

\begin{figure}
	\includegraphics[width=1.0\linewidth]{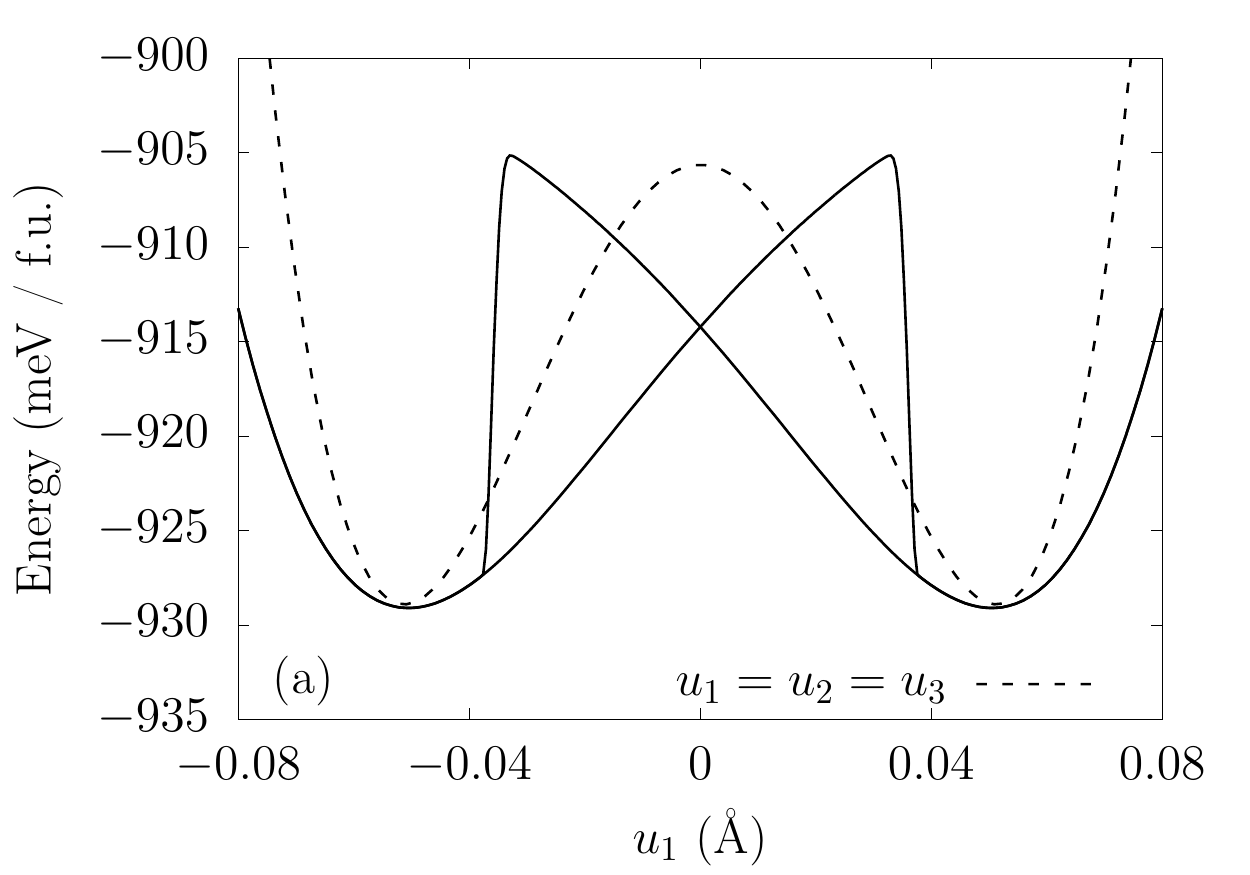}\\
	\includegraphics[width=1.0\linewidth]{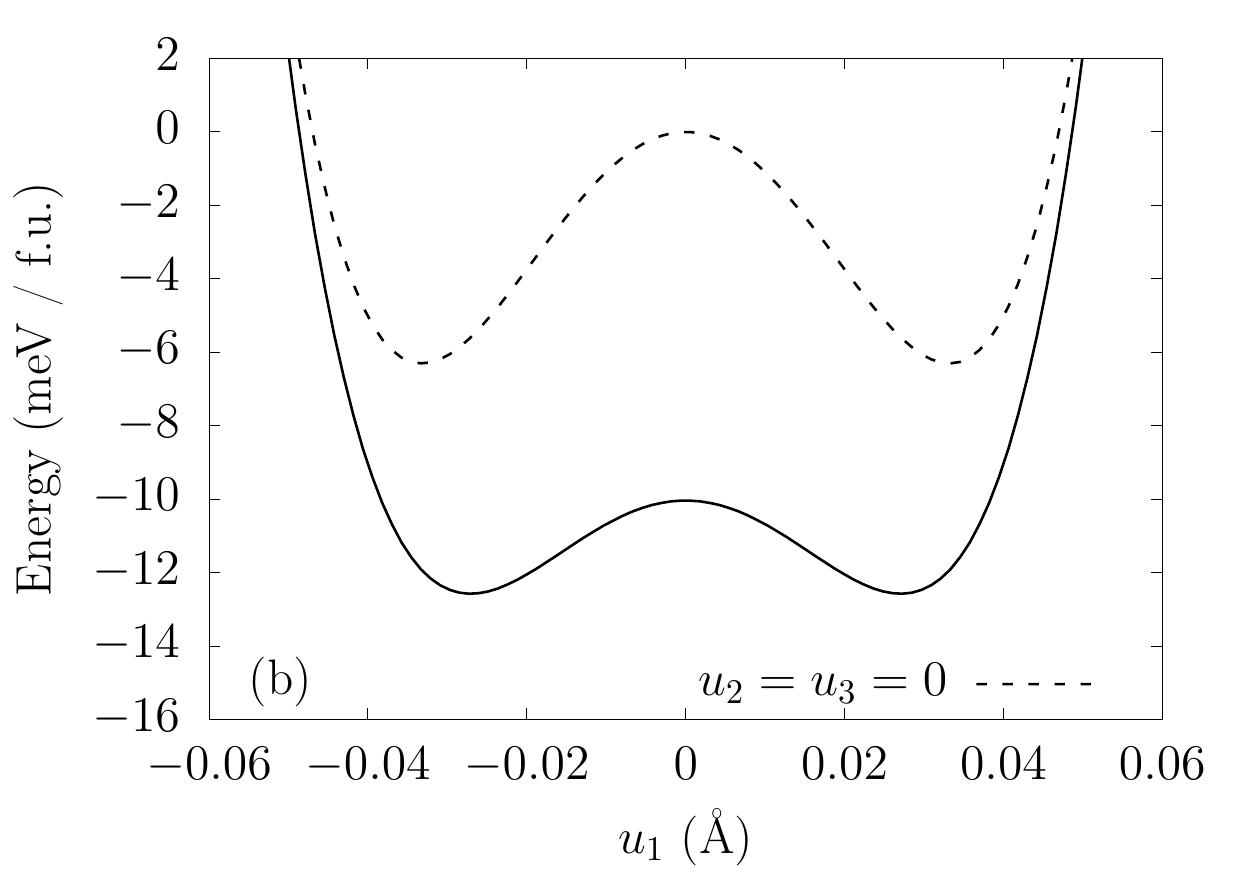}\\
	\caption{Potential energy landscape for LiOsO$_3$ (a) and BaTiO$_3$ (b) 
  from our first principles effective Hamiltonians, obtained by minimizing the energy at fixed $u_1$. For LiOsO$_3$, a double-well like curve is obtained when $u_1=u_2=u_3$ is enforced (dashed line) 
  and a butterfly-like diagram is obtained when all the parameters are allowed to evolve freely (solid line). 
  For BaTiO$_3$, the dashed line represents the study under the $u_2=u_3=0$ constraint, and the solid line the 
  case with no constraints.}
	\label{fig:double_well}
\end{figure}

The resulting double-well potential curves of LiOsO$_3$ and BaTiO$_3$ are shown in Fig. \ref{fig:double_well}.
Before commenting on the results, it is useful to recall the 
structural properties of each of the two materials.
The structural ground state of LiOsO$_3$ has $R3c$ symmetry,
containing both polar distortions and antiphase octahedral tilts 
 ($a^-a^-a^-$ in Glazer notation) oriented along the $[111]$ 
 pseudocubic direction. 
Since the energy scale associated to the AFD distortions is an order
of magnitude larger than that associated to ${\bf u}$, they are unlikely 
to be affected by a weak elastic field; in practice, ${\bf u}$ can only switch 
between the $[111]$ and $[\bar{1}\bar{1}\bar{1}]$ states.
Regarding the actual switching path, two scenarios are in principle possible. If the non-polar $R\bar{3}c$ structure 
were stable under the constraint $u_1=0$, the polar modes would be forced to evolve along the same pseudocubic [111] direction even under the action of a $[100]$-oriented external force. However, previous first-principles calculations have shown \cite{sim2014first} that the $R\bar{3}c$ phase has more 
than one imaginary mode at $\Gamma$, which means that $R\bar{3}c$ is unlikely to be the saddle point.
This suspicion is nicely confirmed by the results of our effective Hamiltonian: indeed, 
the ``butterfly diagram'' of Fig. \ref{fig:double_well} clearly reflects the presence of a switchable in-plane 
polarization at $u_1=0$; the resulting coercive field is
$F_{\text{coerc}}$=0.34 eV/$\text{\AA}$. 
To quantify how much the system gains by circumnavigating the energy barrier, we attempted the same computational experiment while imposing $u_1=u_2=u_3$ along the path; as expected, we obtain a substantially larger critical field of $F_{\text{coerc}}$=0.69 eV/$\text{\AA}$, assuming that the field is still applied along [100]. %

For BaTiO$_3$ the polarization cannot be constrained by the tilts, 
since the latter are absent in this material. 
At low-temperature BaTiO$_3$ has $R3m$ symmetry, and we find that
the lowest switching barrier occurs when the polarization
continuously rotates from [111] to [$\bar{1}11$] by passing through an
orthorhombic [110] saddle point. (The path is roughly 
oriented along [100]). We find a critical coercive field of 
$F_{\text{coerc}}$=0.14 eV/$\text{\AA}$ for such a switching path.
For comparison to room-temperature experiments, where BaTiO$_3$ adopts
a tetragonal structure, we also calculate 
the hypothetical barrier that one would obtain by constraining 
${\bf P \parallel}$ [100] (i.e. by setting the in-plane components of ${\bf P}$
to zero). We find $F_{\text{coerc}}$=0.29 eV/$\text{\AA}$. This is a substantially
larger value than the aforementioned threshold for polarization rotation, in line with 
literature results.

We are now ready to answer  the main physical question we asked to ourselves at the beginning: how much do we need to bend a LiOsO$_3$ sample to reverse its polar lattice distortion?
By means of Eq. (\ref{eq:R_crit}) we can compute the critical bending radius for both materials. 
The obtained values are $R_{\rm crit}\sim 118$ $\text{\AA}$ for LiOsO$_3$ and 
$R_{\rm crit}\sim 125$ $\text{\AA}$ for rhombohedral BaTiO$_3$ and $R_{\rm crit}\sim 60$ $\text{\AA}$ for tetragonal BaTiO$_3$. 
Remarkably, the calculated critical bending radius of LiOsO$_3$ is twice as large as that of tetragonal BaTiO$_3$, essentially matching the calculated value of rhombohedral BaTiO$_3$.
Since mechanical switching of polar domains in tetragonal barium titanate has already been experimentally achieved \cite{ovcenavsek2015nanomechanics} via strain gradients, our results indicate that this is very likely to be feasible in LiOsO$_3$ as well.

Our results for $\text{BaTiO}_3$ are in good agreement with the ones reported in Ref. \cite{li2017shear} 
where a critical bending radius of 110 $\text{\AA}$ was estimated. 
This is substantially smaller than the available experimental estimates 
(a value of  $R_{\rm crit}\sim300$ $\text{\AA}$ was observed in 
 $\text{BaTiO}_3$~\cite{ovcenavsek2015nanomechanics}).
This is expected: theoretical estimations of coercive fields in ferroelectrics that are based on the homogeneous Landau potential
are typically overestimated by one or two orders of magnitude~\cite{shelke2011reduced}.
Consideration of more realistic mechanisms (e.g., domain wall nucleation and motion) would drastically complicate our study, and bring us far from our main scopes. We stress in any case, that our underestimation of the critical bending radii compared to experiments  should be ascribed to an overestimation of $F_{\rm crit}$, while we regard our calculation of the flexocouplings as accurate. (The contribution of the oxygen octahedral tilt gradients to the flexocoupling in lithium osmate was neglected in this work. While certainly present, we consider it unlikely to qualitatively affect our conclusions; it will be an interesting topic for follow-up studies.)

We expect our results to significantly broaden the scopes of both flexoelectricity and the ongoing search for new functionalities based on polar metals. In this sense, we believe that our work may open several unexplored research directions. First and foremost, we regard an experimental verification of our predictions as the most pressing priority. Second, it will be interesting to estimate the magnitude of the flexocouplings in a broader range of polar metals, and identify candidates where the effect is especially strong. Finally, from the point of view of the theory, developing the methodological tools to assess the impact of tilt gradients on the calculated coefficients is another topic that we regard as promising for future studies.

\begin{acknowledgments}
We acknowledge the support of Ministerio de Economia,
Industria y Competitividad (MINECO-Spain) through
Grants  No.  MAT2016-77100-C2-2-P  and  No.  SEV-2015-0496,
and  of Generalitat de Catalunya (Grant No. 2017 SGR1506).
This project has received funding from the European
Research Council (ERC) under the European Union's
Horizon 2020 research and innovation program (Grant
Agreement No. 724529). Part of the calculations were performed at
the Supercomputing Center of Galicia (CESGA).
\end{acknowledgments}

\bibliographystyle{apsrev4-1custom}
\bibliography{biblio.bib}

\end{document}


\beginsupplement

\title{Supplementary notes for "Switching a polar metal via strain gradients"}
\author{Asier Zabalo} 
\affiliation{Institut de Ci\`encia de Materials de Barcelona (ICMAB-CSIC), Campus UAB, 08193 Bellaterra, Spain}
\author{Massimiliano Stengel}
\affiliation{Institut de Ci\`encia de Materials de Barcelona (ICMAB-CSIC), Campus UAB, 08193 Bellaterra, Spain}
\affiliation{ICREA-Instituci\'o Catalana de Recerca i Estudis Avançats, 08010 Barcelona, Spain}

\maketitle

\onecolumngrid

\section{ I.  Computational details}

Calculations on $\text{LiOsO}_3$ are carried out with pseudopotentials under the generalized 
gradient approximation (GGA) with Perdew-Burke-Ernzerhof (PBE) exchange-correlation 
functionals \cite{perdew1996generalized}. 
A dense $16\cross16\cross16$ $\mathbf{k}$-point mesh is used to sample the Brillouin zone 
with a plane-wave cutoff of 60 Ha. 
A Gaussian smearing of 0.01 Ha is applied and effects of spin-orbit coupling are neglected 
\cite{benedek2016ferroelectric}. 
The dynamical properties, i.e., phonon frequencies and eigenmodes, are calculated using  
DFPT as implemented in \textsc{abinit}. 
The independent components of both the elastic and the flexocoupling tensor are computed 
with Eq. (2) and Eq. (6) of the main text and convergence is tested 
with respect to the two-dimensional grid of \textbf{q} points. 
For cubic systems a two dimensional grid is enough to calculate the three independent 
components of the mentioned tensors, commonly denoted as longitudinal ($C_{11},f_{11}$), 
transverse ($C_{12},f_{12}$) and shear ($C_{44},f_{44}$). 
For a $16\cross16\cross1$ mesh of \textbf{q} points convergence was clearly reached for 
the elastic constants, and as a consequence, for the flexocoupling tensor.
(Remember that both calculations are carried out using exactly the same method, so a similar 
convergence rate and errors are expected.) 
Calculations for BaTiO$_3$ are performed by using LDA and a $12\times 12\times 12$ ${\bf k}$-point 
mesh, and the flexocoupling coefficients are extracted via the method described in Ref.~\cite{stengel2016unified}.
For a meaningful comparison with metallic LiOsO$_3$ we use the conduction band bottom as energy
reference, which yields the $n$-type flexocoupling coefficients as defined in Ref.~\cite{stengel2016unified}.
(Note that the $f_{44}$ component is independent of such a choice.)

\section{ II. First-principles effective Hamiltonian of LiOsO$_3$}
\textbf{Expression and parameters} The explicit expression for the first-principles Hamiltonian of LiOsO$_3$, which gives the energy of a 10-atom cell, is the following:
\begin{equation}
\begin{split}
H&=\frac{\Omega}{2}C_{11}(s_1^2+s_2^2+s_3^2) + \Omega C_{12}(s_1s_2+s_1s_3+s_2s_3)+\frac{\Omega}{2}C_{44}(s_4^2+s_5^2+s_6^2)\\
&+\beta_1(q_1^2+q_2^2+q_3^2)+\beta_{2}(q_1^4+q_2^4+q_3^4)+\beta_{3}(q_2^2q_3^2+q_1^2q_3^2+q_1^2q_2^2)\\
&+\zeta_1(u_1^2+u_2^2+u_3^2)+\zeta_2(u_1^4+u_2^4+u_3^4) +\zeta_3(u_2^2u_3^2+u_1^2u_3^2+u_1^2u_2^2)\\
&+\zeta_4(u_1^6+u_2^6+u_3^6) + \zeta_5[u_1^4(u_2^2+u_3^2)+u_2^4(u_1^2+u_3^2)+u_3^4(u_1^2+u_2^2)]\\
&+\zeta_6(u_1^8+u_2^8+u_3^8) \\
&+\lambda_1(s_1q_1^2+s_2q_2^2+s_3q_3^2)+\lambda_2[s_1(q_2^2+q_3^2)+s_2(q_1^2+q_3^2)+s_3(q_1^2+q_2^2)]\\
&+\lambda_3(s_4q_2q_3+s_5q_1q_3+s_6q_1q_2)+\lambda_4(s_1q_1^4+s_2q_2^4+s_3q_3^4)\\
&+\lambda_5(s_4q_2^2q_3^2+s_5q_1^2q_3^2+s_6q_1^2q_2^2)\\
&+\rho_1(s_1u_1^2+s_2u_2^2+s_3u_3^2)+\rho_2[s_1(u_2^2+u_3^2)+s_2(u_1^2+u_3^2)+s_3(u_1^2+u_2^2)]\\
&+\rho_3(s_4u_2u_3+s_5u_1u_3+s_6u_1u_2)+\rho_4(s_1u_1^4+s_2u_2^4+s_3u_3^4)\\
&+\rho_5(s_4u_2^2u_3^2+s_5u_1^2u_3^2+s_6u_1^2u_2^2)\\
&+\rho_6[s_1(u_2^4+u_3^4)+s_2(u_1^4+u_3^4)+s_3(u_1^4+u_2^4)]\\
&+\gamma_1(u_1^2q_1^2+u_2^2q_2^2+u_3^2q_3^2)+\gamma_2(u_1^2q_1^4+u_2^2q_2^4+u_3^2q_3^4).
\end{split}
\end{equation}
where the strain is given in the matrix Voigt notation. The calculated model parameters are given in Table \ref{tab:Ham_parameters}.
\begin{table}[b!]
\caption{Calculated model parameters for the first-principles Hamiltonian of LiOsO$_3$. \label{tab:Ham_parameters}}
\begin{ruledtabular}
\begin{tabular}{ccc|ccc}
Parameter&Value & Units&Parameter&Value&Units\\
\hline
$\Omega$ & 112.94 & $\text{\AA}^3$& $\lambda_1$ & 38.2516 & eV/$\text{\AA}^2$ \\
$C_{11}$ & 365.58 & GPa& $\lambda_2$ & 19.0951 & eV/$\text{\AA}^2$ \\
$C_{12}$ & 129.48 & GPa& $\lambda_3$ & -12.3815 & eV/$\text{\AA}^2$ \\
$C_{44}$ & 44.09 & GPa&  $\lambda_4$ & -32.8244 & eV/$\text{\AA}^4$ \\
$\beta_{1}$ & -4.31776  & eV/$\text{\AA}^2$& $\lambda_5$ & 16.3332 & eV/$\text{\AA}^4$ \\
$\beta_{2}$ & 8.03626 & eV/$\text{\AA}^4$& $\rho_1$ & 81.4362 & eV/$\text{\AA}^2$ \\
$\beta_{3}$ & 5.97264 & eV/$\text{\AA}^4$& $\rho_2$ & -136.627 & eV/$\text{\AA}^2$ \\
$\zeta_{1}$ & -35.4382 & eV/$\text{\AA}^2$& $\rho_3$& -110.971 & eV/$\text{\AA}^2$ \\
$\zeta_{2}$ & 868.601 & eV/$\text{\AA}^4$& $\rho_4$& -308.951 &  eV/$\text{\AA}^4$\\
$\zeta_{3}$ & 2205.6 & eV/$\text{\AA}^4$& $\rho_5$ & -14380.5 & eV/$\text{\AA}^4$\\
$\zeta_{4}$ & -6978.34 & eV/$\text{\AA}^6$& $\rho_6$ & -11808.3 & eV/$\text{\AA}^4$ \\
$\zeta_{5}$ & 59153.0 & eV/$\text{\AA}^6$& $\gamma_1$& 348.113 &  eV/$\text{\AA}^4$\\
$\zeta_{6}$ & 20643.6 & eV/$\text{\AA}^8$& $\gamma_2$ & -990.2  &eV/$\text{\AA}^6$ 
\end{tabular}
\end{ruledtabular}
\end{table}

\clearpage

\begin{figure}[b!]
 	\includegraphics[width=0.6\linewidth]{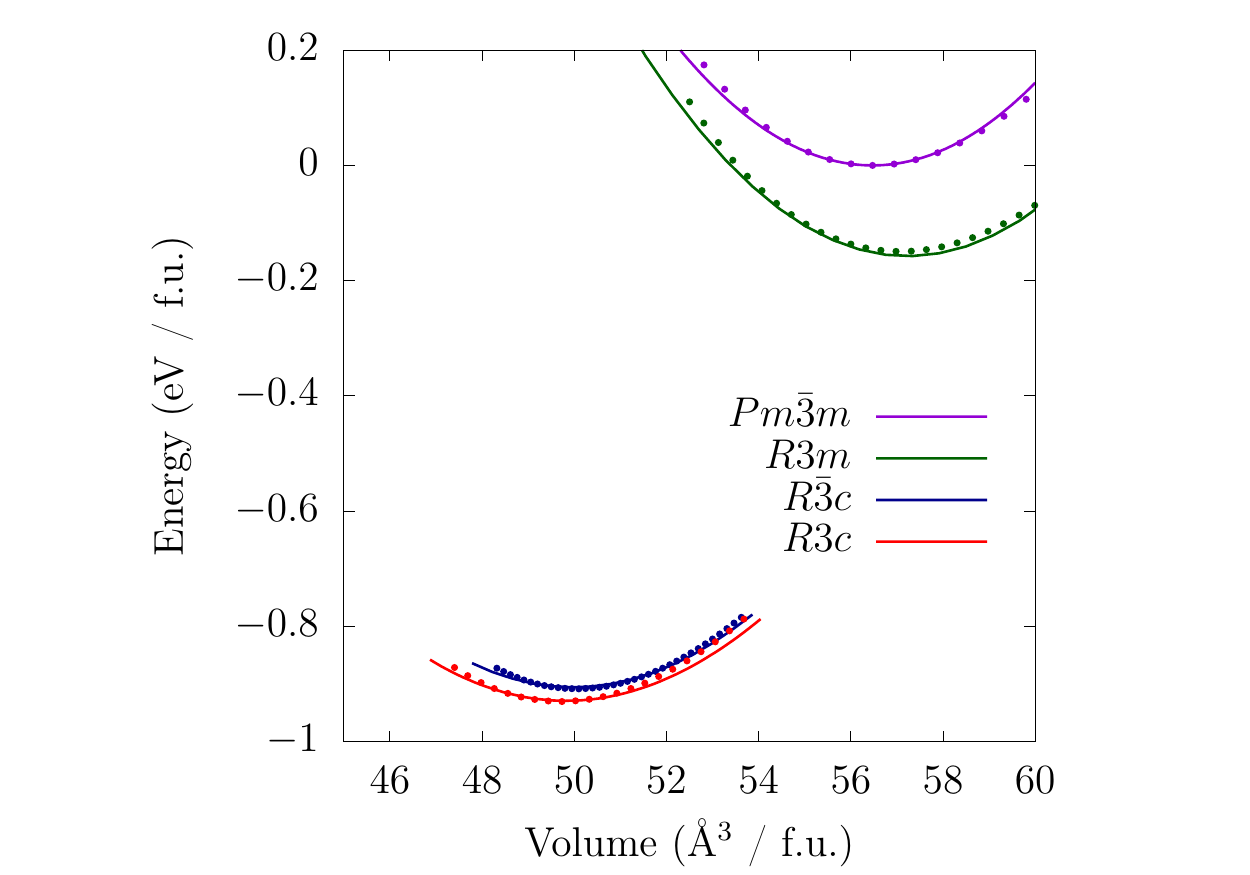}
 	\caption{Energy per formula unit as a function of the volume of the perovskite formula unit of $Pm\bar{3}m$, $R3m$, $R\bar{3}c$, and $R3c$ phases. Solid lines are obtained from our effective Hamiltonian.
 		\label{fig:energies_phases}}
 \end{figure}

\textbf{Validation of the model} 
 As a first test of the model we compute the energy 
 of the most relevant phases that can be obtained by combining 
 antiferrodistortive and polar instabilities, compared with the
 reference $Pm\bar{3}m$ structure.
 This is, of course, the structural ground state of $R3c$ symmetry,
 containing both polar distortions and antiphase octahedral tilts 
 ($a^-a^-a^-$ in Glazer notation) oriented along the $[111]$ 
 pseudocubic direction.
 Other important phases are the centrosymmetric $R\bar{3}c$
 and the polar $R3m$, where either the polar or the 
 antiferrodistortive modes are suppressed.
 As we can see from the calculated values (Table \ref{tab:DFT_model}), 
 the model accurately reproduces the first-principles results 
 obtained by fully relaxing the crystal cell within the 
 given symmetry.
\begin{table}[b!]
 	\caption{Volume and energy comparison between first-principles calculations and our model. Energies are in meV /.f.u. units.
 		\label{tab:DFT_model}}
	\begin{ruledtabular}
 		\begin{tabular}{ccc|cc}
 			& \multicolumn{2}{c}{DFT} &  \multicolumn{2}{c}{Effective Hamiltonian}\\
 			\hline
 			&  $\Delta E$ & $\Delta V$(\%)   & $\Delta E$ & $\Delta V$(\%) \\
 			$R3m$ & -149 & 1.1 & -157 & 1.3\\
 			$R\bar{3}c$ & -908 & -11.4 & -906 & -11.5\\
 			$R3c$ &  -930  & -12.0 & -929   & -11.9 \\
 		\end{tabular}
 	\end{ruledtabular}
\end{table}
 In particular, this first-principles Hamiltonian almost perfectly 
 describes the energy difference between the lowest-energy $R\bar{3}c$ and the $R3c$ phases, which are the most relevant for ferroelectric switching: 
 -22 meV/f.u. with DFT and -23 meV/f.u. with our Hamiltonian, were f.u. 
 stands for perovskite formula unit. As a further test, we study the phase diagram of LiOsO$_3$ as a function of cell volume.
 The dots shown in Fig. \ref{fig:energies_phases} were obtained 
 via explicit DFT calculations
 by fully relaxing the ionic positions 
 for each fixed cell volume. 
 These results are in excellent agreement with experimental values \cite{shi2013ferroelectric} 
 and previous first-principles studies \cite{sim2014first}. 
 Solid lines represent the energies obtained with our model, again 
 showing excellent agreement with the first-principles values. 
%
%
\section{ III.  First-principles effective Hamiltonian of $\text{BaTiO}_3$}
\textbf{Expression and parameters}
The explicit expression for the first-principles Hamiltonian of BaTiO$_3$, which gives the energy of a 5-atom cell, is the following:
\begin{equation}
\begin{split}
H&=\frac{\Omega}{2}C_{11}(s_1^2+s_2^2+s_3^2) + \Omega C_{12}(s_1s_2+s_1s_3+s_2s_3)+\frac{\Omega}{2}C_{44}(s_4^2+s_5^2+s_6^2)\\
&+\zeta_1(u_1^2+u_2^2+u_3^2) + \zeta_2(u_1^4+u_2^4+u_3^4) +\zeta_3(u_2^2u_3^2+u_1^2u_3^2+u_1^2u_2^2)\\
&+\rho_1(s_1u_1^2+s_2u_2^2+s_3u_3^2)+\rho_2[s_1(u_2^2+u_3^2)+s_2(u_1^2+u_3^2)+s_3(u_1^2+u_2^2)]\\
&+\rho_3(s_4u_2u_3+s_5u_1u_3+s_6u_1u_2)+\rho_4(s_1u_1^4+s_2u_2^4+s_3u_3^4)\\
&+\rho_5(s_4u_2^2u_3^2+s_5u_1^2u_3^2+s_6u_1^2u_2^2)\\
&+\rho_6[s_1(u_2^4+u_3^4)+s_2(u_1^4+u_3^4)+s_3(u_1^4+u_2^4)]
\end{split}
\end{equation}
The calculated model parameters are given in Table \ref{tab:Ham_parameters_BaTiO3}.

\begin{table}[h!]
	\caption{Calculated model parameters for the first-principles Hamiltonian of $\text{BaTiO}_3$. \label{tab:Ham_parameters_BaTiO3}}
	\begin{ruledtabular}
		\begin{tabular}{ccc|ccc}
			Parameter&Value & Units&Parameter&Value&Units\\
			\hline
			$\Omega$ & 60.745 & $\text{\AA}^3$& $\rho_1$ & -911.228& eV/$\text{\AA}^2$\\
			$C_{11}$ & 353.30 & GPa& $\rho_2$ & -97.0643 & eV/$\text{\AA}^2$ \\
			$C_{12}$ & 121.75 & GPa& $\rho_3$ & -99.2179 & eV/$\text{\AA}^2$\\
			$C_{44}$ & 137.72 & GPa& $\rho_4$ & -29262.0 & eV/$\text{\AA}^4$\\
			$\zeta_1$ & -11.5402& eV/$\text{\AA}^2$& $\rho_5$ & 31131.0 & eV/$\text{\AA}^4$  \\
			$\zeta_2$ & 7294.25 & eV/$\text{\AA}^4$& $\rho_6$ & 9691.01 & eV/$\text{\AA}^4$ \\
			$\zeta_3$ & 3441.25 & eV/$\text{\AA}^4$ & --- & --- &--- 
			
		\end{tabular}
	\end{ruledtabular}
\end{table}

\begin{figure}[b!]
	\includegraphics[width=0.6\linewidth]{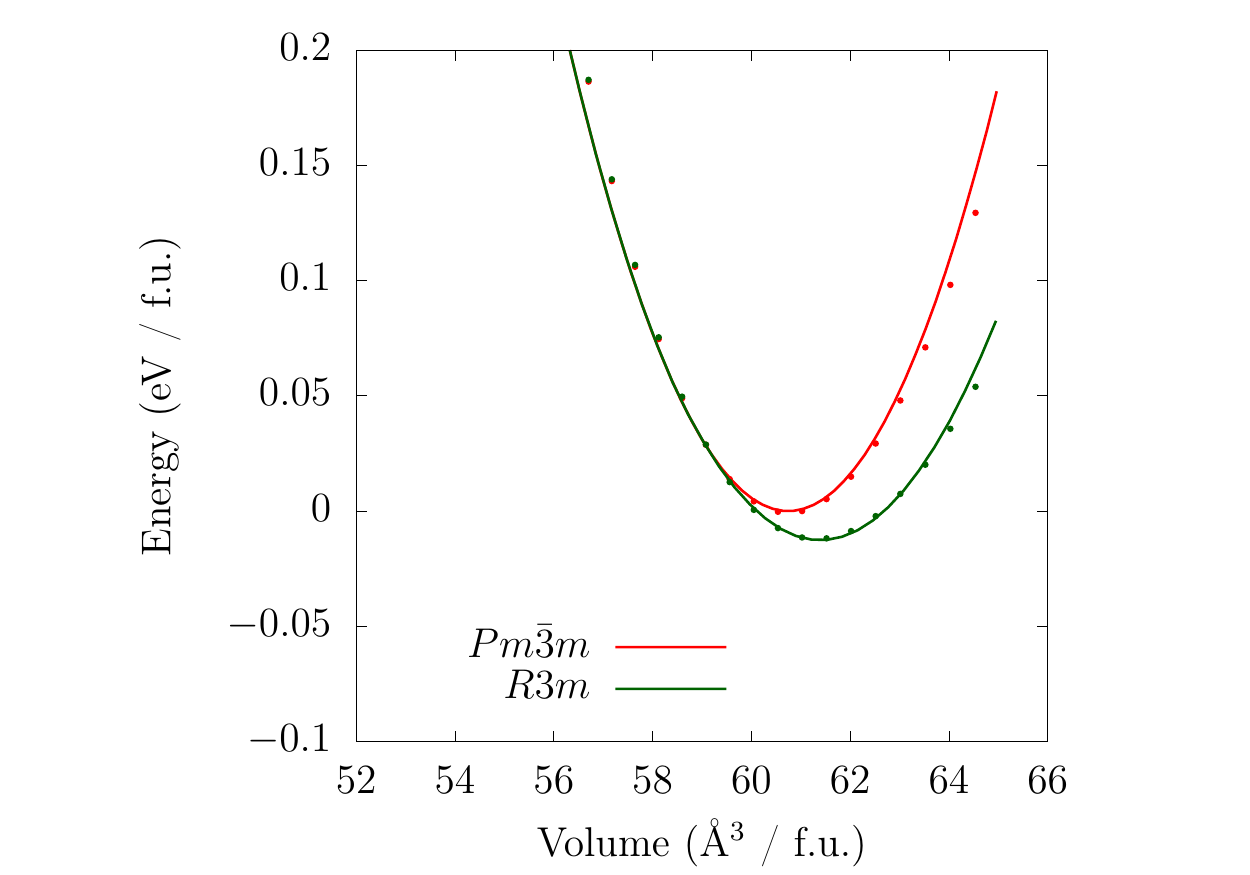}
	\caption{Energy per formula unit as a function of the volume of the perovskite formula unit of $Pm\bar{3}m$ and $R3m$ phases of BaTiO$_3$. Solid lines are obtained from our effective Hamiltonian.
		\label{fig:BaTiO3_curves}}
\end{figure}

\textbf{Validation of the model} As with lithium osmate, we first compute the energy of the relevant phases. Since there are no tilts, this reduces to calculating the energy of the ground state $R3m$ phase. Fig. \ref{fig:BaTiO3_curves} shows the phase diagram of the cubic and $R3m$ phases as a function of cell volume, showing an excellent agreement with the first principles calculations.

\clearpage

\section{ IV.  Flexocoupling coefficients}
\begin{figure}[h!]
	\includegraphics[width=0.49\linewidth]{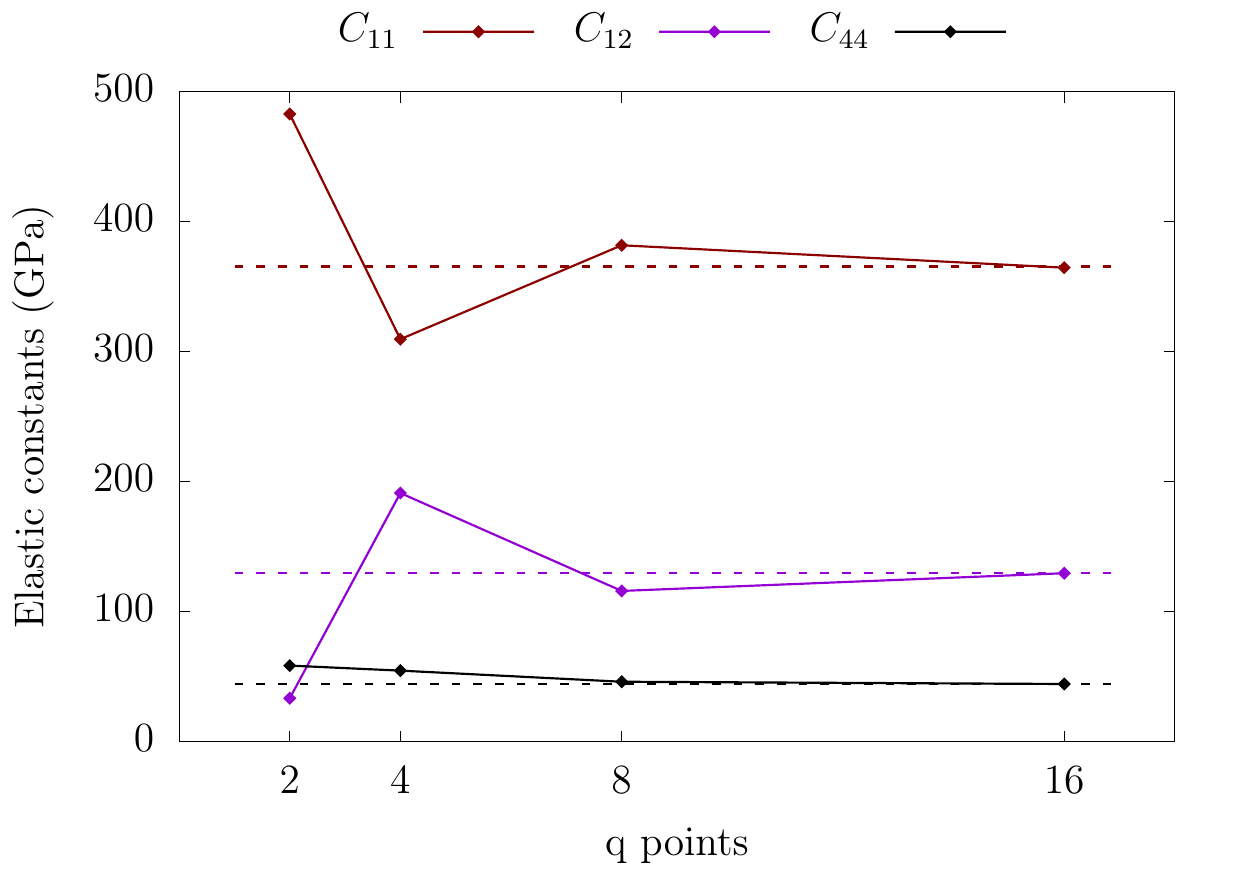}
	\includegraphics[width=0.49\linewidth]{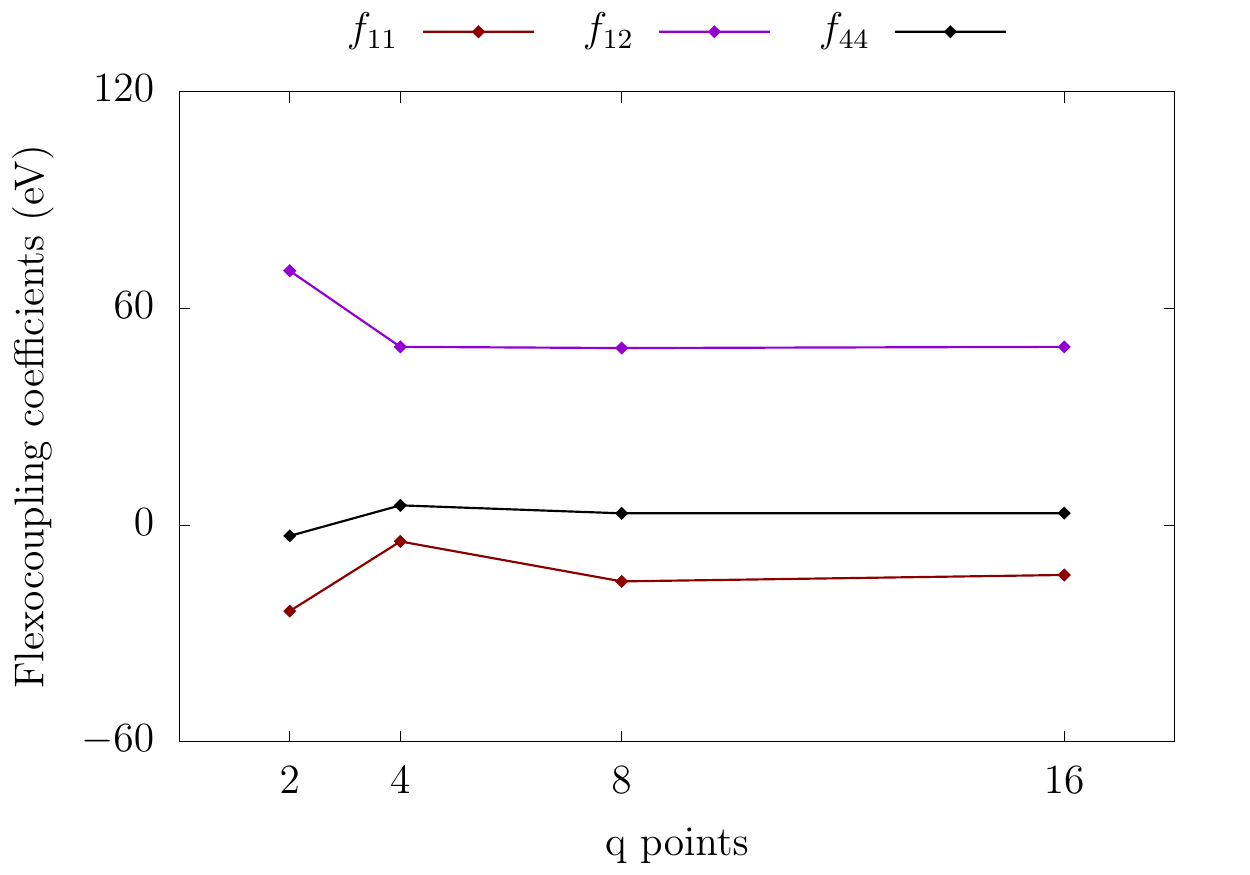}\\
	\caption{Convergence of the elastic tensor and the flexocoupling tensor with respect to the two-dimensional grid of q points ($q\cross q\cross 1$). The dashed lines represent the reference values of the elastic constants, computed with DFPT. Only the three independent components are shown. \label{fig:elastic_flexo_q}}
\end{figure}

\textbf{Convergence of flexocoupling coefficients} The convergence of the flexocoupling coefficients (in the type-II representation) of LiOsO$_3$ with respect to the \textbf{q} point mesh, for a fixed \textbf{k} point mesh, is analyzed.  Numerical values are shown in Table \ref{tab:elastic_flexo_q} and graphical results are given in Fig. \ref{fig:elastic_flexo_q}.
\begin{table}[h!]
	\caption{Independent components of the elastic tensor (in GPa) and the flexocoupling tensor (in eV). \label{tab:elastic_flexo_q}}
	\begin{ruledtabular}
		\begin{tabular}{ccccc}
			grid of \textbf{q} points & $2\cross2\cross1$& $4\cross4\cross1$ & $8\cross8\cross1$& $16\cross16\cross1$\\
			\hline
			$C_{11}$& 482.93 &309.66 & 381.90 & 364.72\\
			$C_{12}$& 33.45 & 191.35  & 115.92& 129.54\\
			$C_{44}$& 58.46  &54.63  & 46.13& 44.30\\
			\hline
			$f^{\text{II}}_{11}$& -23.84 & -4.53 & -15.62 & -13.82 \\
			$f^{\text{II}}_{12}$& 70.42 & 49.34 & 49.02 & 49.35 \\
			$f^{\text{II}}_{44}$& -3.43 & 5.43 & 3.24 & 3.27 \\
		\end{tabular}
	\end{ruledtabular}
\end{table}

\textbf{Effective flexocoupling coefficients} In order to have a self contained manuscript, here we give the expressions of the effective flexocoupling coefficients for the [100], [110] and [111] orientations for the beam-bending limit following the definitions of Ref. \cite{narvaez2015large}.\\
\begin{equation}
\begin{split}
f_{\text{eff}}^{100}&=\frac{-C_{12}}{C_{11}+C_{12}}f^{\text{II}}_{11}+\frac{C_{11}}{C_{11}+C_{12}}f^{\text{II}}_{12},\\
f_{\text{eff}}^{110}&=Af^{\text{II}}_{11}+Bf^{\text{II}}_{12}-2(1-A)f_{44}^{\text{II}},\\
f_{\text{eff}}^{111}&=\frac{C_{44}}{C_{11}+2C_{12}+C_{44}}(f_{11}^{\text{II}}+2f_{12}^{\text{II}})-\frac{C_11+2C_{12}}{C_{11}+2C_{12}+C_{44}}f_{44}^{\text{II}},
\end{split}
\end{equation}
where
\begin{equation}
\begin{split}
A&=\frac{2C_{11}C_{44}}{(C_{11}-C_{12})(C_{11}+2C_{12})+2C_{11}C_{44}},\\
B&=\frac{2(C_{11}-2C_{12})C_{44}}{(C_{11}-C_{12})(C_{11}+2C_{12})+2C_{11}C_{44}}.
\end{split}
\end{equation}

%
